\documentclass[twocolumn,aps,prl,preprintnumbers,bibnotes10pt,superscriptaddress]{revtex4}

\usepackage{amsmath}
\usepackage{graphicx}
\usepackage{ulem}
\usepackage{dcolumn}
\usepackage{epsfig}
\usepackage{bm}
\usepackage{array}
\usepackage[frenchb]{babel}
\usepackage{hyperref}
\hypersetup{
colorlinks=true,
citecolor=black,
linkcolor=red,
urlcolor=black
 , bookmarks=true,
 pdfmenubar=true
}

\usepackage{hyperref}
\usepackage{graphicx}
\usepackage{amsmath}
\usepackage{amsfonts}
\usepackage{amssymb}
\usepackage{xcolor}
\usepackage{bbm}
\usepackage{calrsfs}
\usepackage{dutchcal}
\usepackage{amsthm}

\newcommand{\bra}[1]{\ensuremath{\langle#1|}}
\newcommand{\ket}[1]{\ensuremath{|#1\rangle}}
\newcommand{\Eins}{\ensuremath{\mathbbm 1}}
\newcommand{\mean}[1]{\ensuremath{\big\langle #1 \big\rangle}}

\newcommand{\vect}[1]{\bm{#1}}

\newcommand{\be}{\begin{equation}}
\newcommand{\ee}{\end{equation}}

\newcommand{\beq}{\begin{eqnarray}}
\newcommand{\eeq}{\end{eqnarray}}

\newcommand{\J}{\hat{J}}

\begin{document}

\title{Quantum Phase Estimation Algorithm with Gaussian Spin States}
\author{Luca Pezz\`e} 
\affiliation{QSTAR, INO-CNR and LENS, Largo Enrico Fermi 2, 50125 Firenze, Italy} 

\author{Augusto Smerzi}
\affiliation{QSTAR, INO-CNR and LENS, Largo Enrico Fermi 2, 50125 Firenze, Italy} 

\begin{abstract}
Quantum phase estimation (QPE) is one of the most important subroutines in quantum computing. 
In general applications, current QPE algorithms either suffer an exponential time overload
or require a set of -- notoriously quite fragile -- GHZ states.
These limitations have prevented so far the demonstration of QPE beyond proof-of-principles. 
Here we propose a new QPE algorithm that scales linearly with time 
and is implemented with a cascade of Gaussian spin states (GSS).
GSS are renownedly resilient and have been created experimentally in a variety of platforms, from hundreds of ions up to millions of cold/ultracold neutral atoms. 
We show that our protocol
achieves a QPE sensitivity overcoming previous bounds, including those obtained with GHZ states, 
and is robustly resistant to several sources of noise and decoherence. 
Our work paves the way toward realistic quantum advantage demonstrations of the QPE, as well as applications of atomic squeezed states for quantum computation.
 \end{abstract}

\maketitle
\date{\today}

\section {Introduction}

Quantum phase estimation (QPE)~\cite{Kitaev, KitaevBOOK, ClevePRSA1998} is the building block of known quantum computing algorithms 
providing exponential speedup~\cite{NielsenBOOK}, including the computation of the
eigenvalues of Hermitian operators~\cite{AbramsPRL1999}, such as molecular spectra~\cite{Aspuru-GuzikSCIENCE2005, McArdleRMP2020}, 
number factoring~\cite{ShotSIAM1997, Martin-LopezNATPHOT2012, MonzSCIENCE2016},
and quantum sampling~\cite{TemmeNATURE2011}.
All these applications require the calculation of an eigenvalue of a unitary matrix
$U \ket{u} = e^{-i \theta}\ket{u}$, where $\ket{u}$ is 
the corresponding eigenstate, which can be cast as the estimation of an unknown phase $\theta \in [-\pi, \pi)$.
The information about $\theta$ is encoded into one or more ancilla qubits via multiple applications of a controlled-$U$ gate~\cite{NielsenBOOK}.
The QPE problem plays a key role
also in the alignment of spatial reference frames \cite{RudolphPRL2003} and clock synchronisation~\cite{DeBurghPRA2005}, with further developments in  
atomic clocks \cite{KesslerPRL2014}, and worldwide networks \cite{KomarNATPHY2014}.

According to the current paradigm~\cite{Kitaev, KitaevBOOK}, QPE algorithms are implemented iteratively, without requiring 
the inverse quantum Fourier transform~\cite{ClevePRSA1998, NielsenBOOK, GriffithsPRL1996, vanDamPRL2007}. 
Iterative QPE consists of multiple steps, each step being realized in two possible ways:
i) Using a single ancilla qubit $(\ket{0} + \ket{1})/\sqrt{2}$ that interrogates $2^k$ times the controlled-$U$ gate in 
temporal sequence~\cite{HigginsNATURE2007, PaesaniPRL2017, BonatoNATNANO2016}.
In this way, the ancilla qubit is transformed to $(\ket{0} + e^{i 2^k \theta} \ket{1})/\sqrt{2}$ and
the phase information is extracted via a Hadamard gate $H$, followed by a projection in the computational basis.
ii) Using $2^k$ ancilla qubits in a GHZ state $(\ket{0}^{\otimes 2^k} + \ket{1}^{\otimes 2^k})/\sqrt{2}$ that interrogate the controlled-$U$ gate 
in parallel~\cite{MitchelSPIE2005, PezzeEPL2007, BerryPRA2009, KesslerPRL2014, KomarNATPHY2014}.
In this case, the GHZ state is transformed to $(\ket{0}^{\otimes 2^k} + e^{i 2^k \theta} \ket{1}^{\otimes 2^k})/\sqrt{2}$ and 
the phase can be extracted by applying a collective Hadamard gate $H^{\otimes N}$ followed by a parity measurement~\cite{MitchelSPIE2005, PezzeEPL2007, BerryPRA2009}.
While the above output states are characterized by a periodicity $2\pi/2^k$ in $\theta$, 
an unambiguous estimate of $\theta \in [-\pi, \pi)$ is obtained by taking 
a sequence of steps using $k = 0, ..., K$ and eventually repeating the procedure $\nu$ times~\cite{Kitaev,KitaevBOOK}.
Using total resources $N_T = \nu \sum_{k=0}^{K} 2^k = \nu ( 2^{K+1} -1)$, it is possible to estimate an unknown $\theta$ 
with a sensitivity reaching the Heisenberg scaling $\Delta \theta \sim 1/N_T$~\cite{HigginsNATURE2007, BerryPRA2009, WiebePRL2016}. 

%%%%%%%%%%%%%%%%%%%%%%%%%%%%%%%%%%%%%%%%%%%%%%%%%%%%%%%%%%%%%%%%%%%%%%%%%%%%%%
%% Figure 1
%%%%%%%%%%%%%%%%%%%%%%%%%%%%%%%%%%%%%%%%%%%%%%%%%%%%%%%%%%%%%%%%%%%%%%%%%%%%%%
\begin{figure*} [t!]
\centering
\includegraphics[width=2\columnwidth]{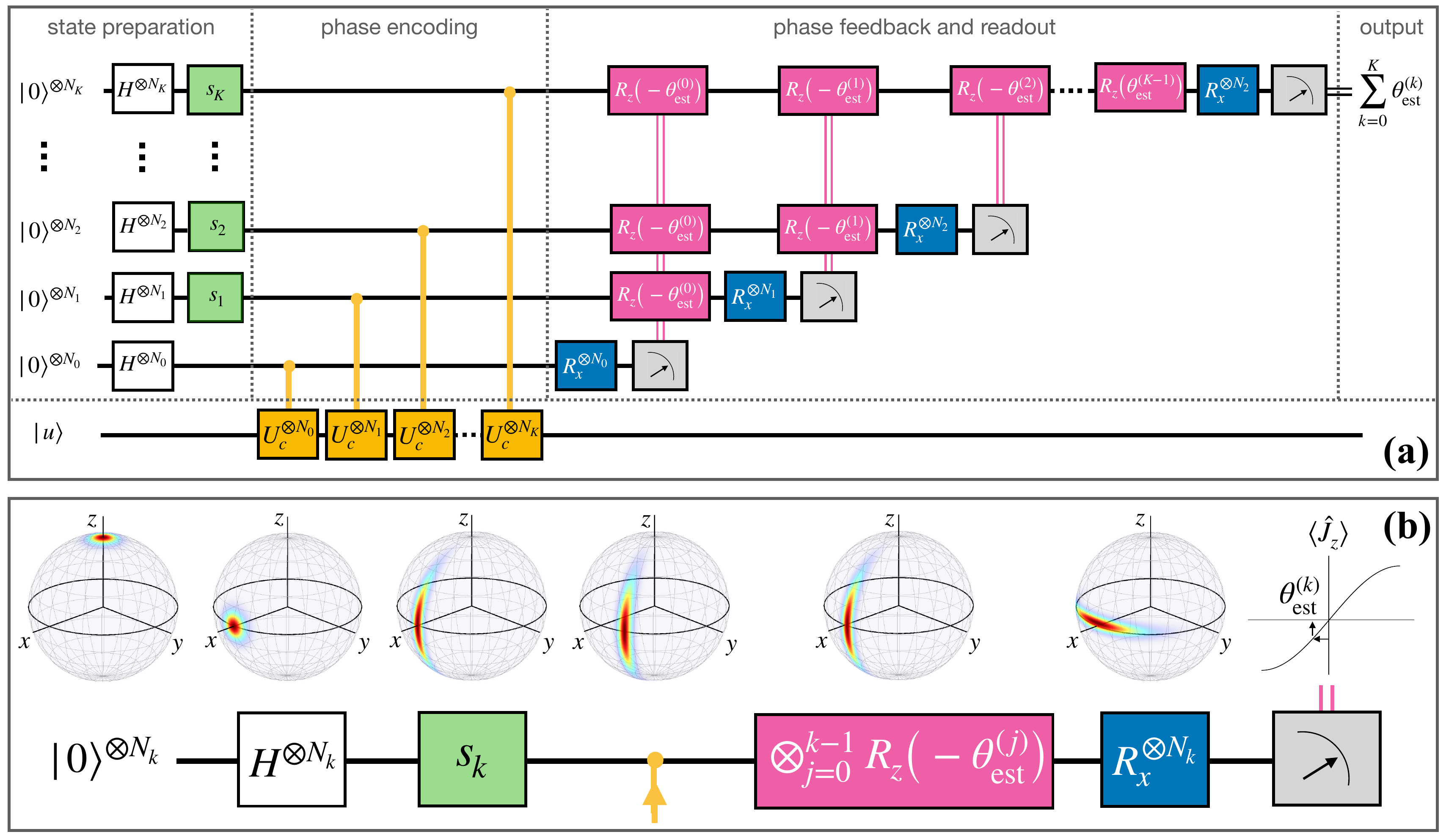}
\caption{{\bf Circuit diagram of quantum phase estimation
algorithm described in this work.} 
(a) The $K+1$ ancilla states prepared in $\ket{0}^{\otimes N_k}$ ($k=0, ..., K$) are first 
rotated by applying collective Hadamard gate $H^{\otimes N_k}$ (white box).
The green box is a squeezing gate, $s_k$ being the squeezing parameter (see text and Methods).
The yellow box is a controlled-$U^{\otimes N_k}$ gate that provides phase encoding conditioned by the eigenstate $\ket{u}$ of the register.
Pink boxes are rotation phase gates $R_z(-\theta_{\rm est}^j)$ implementing a classical phase feedback (pink lines)
based on the estimated $\theta_{\rm est}^{(j)}$ at step $j<k$.
The readout is obtained by applying a rotation gate $R_x^{\otimes N}$ (blue box) followed by a measurement of the eigenstates of $\J_z$ (grey box). 
The output of the algorithm, after $K+1$ steps, is an estimate, $\theta_{{\rm est},K} = \sum_{k=0}^K \theta_{\rm est}^{(k)}$, of the unknown phase $\theta$.
(b) Illustration of the $k$th step of the algorithm (bottom) and the corresponding Husimi representation of the quantum state on the 
$N_k$-qubits Bloch sphere (top) for each operation.
The output signal has a sinusoidal dependence on $\theta$, from which the estimated $\theta_{\rm est}^{(k)}$
is extracted and used to phase feedback the next ($k+1$) step of the algorithm.}
\label{fig1}
\end{figure*} 
%%%%%%%%%%%%%%%%%%%%%%%%%%%%%%%%%%%%%%%%%%%%%%%%%%%%%%%%%%%%%%%%%%%%%%%%%%%%%%
%%%%%%%%%%%%%%%%%%%%%%%%%%%%%%%%%%%%%%%%%%%%%%%%%%%%%%%%%%%%%%%%%%%%%%%%%%%%%%
%%%%%%%%%%%%%%%%%%%%%%%%%%%%%%%%%%%%%%%%%%%%%%%%%%%%%%%%%%%%%%%%%%%%%%%%%%%%%% 

These protocols have critical drawbacks that have precluded the experimental demonstration of quantum advantage. 
The sequential protocol i) considers $2^k$ applications of the controlled-$U$ gate that require exponentially large implementation times. 
This approach has been demonstrated experimentally \cite{HigginsNATURE2007} using 
multiple passes of a single photon through a phase shifter
and it has been more recently applied to eigenvalue estimation~\cite{PaesaniPRL2017} and magnetic field sensing~\cite{BonatoNATNANO2016}.  
The parallel protocol ii) scales linearly with the implementation time but requires 
GHZ maximally entangled states containing exponentially large, $2^k$, number of particles. 
These states are notoriously fragile, since the loss of
even a single particle completely decoheres the state~\cite{EscherNATPHYS2011, DemkowiczNATCOMM2012} and irremediably breaks-down the quantum algorithm. 
GHZ states are currently implemented with up to $20$ particles~\cite{MonzPRL2011, OmranSCIENCE2019}.
Generally speaking, the current {\it grand challenge} in quantum technologies 
is to go beyond proof-of-principle 
demonstrations toward implementations overcoming ``classical" bounds.
In the contest of QPE, it is evident that present shortcomings
call for a radically new approach 
that is, both, scalable with respect to
temporal resources and exploits experimentally robust, easily accessible, quantum states. 

Here we develop a new QPE algorithm that scales linearly with time and 
is implemented with Gaussian spin-squeezed states (GSS)~\cite{SorensenNATURE2001, PezzeRMP2018, MaPR2011}.
The circuit diagram of the quantum algorithm is shown in Fig.~\ref{fig1} and is discussed in details below.
The iterative algorithm is broken ito multiple steps, each implemented with GSS.
GSS have a Gaussian particle statistical distribution and are way more robust to noise and decoherence than GHZ states. Indeed, GSS
have been experimentally demonstrated in a variety of platforms~\cite{PezzeRMP2018}, with a few hundreds ions~\cite{BohnetSCIENCE2016},
several thousands degenerate gases~\cite{GrossNATURE2010, RiedelNATURE2010}, up to 
millions of cold~\cite{LerouxPRL2010, AppelPNAS2009, BohnetNATPHOT2014, HostenNATURE2016} atoms. 
The shortcoming of GSS -- that has prevented so far their use in QPE  -- is that they 
provide high sensitivity only in a relatively small phase interval centred around the
optimal phase value $\theta=0$~\cite{PezzeRMP2018, MaPR2011}.
Our algorithm overcomes this limitation by a classical adaptive phase feedback.
It progressively drives the unknown phase to the optimal sensitivity points of a cascade of GSS that are increasingly squeezed. 
An analytical optimization of the cascade, with respect of both the number of particles and the squeezing parameter of each state, demonstrates a phase sensitivity 
\be \label{eq.HL}
\Delta \theta = \frac{4}{N_T},
\ee 
for the estimation of any arbitrary phase $\theta \in [-\pi, \pi)$. 
This result overcomes the sensitivity obtained with other QPE algorithms proposed so far in the literature~\cite{HigginsNATURE2007, BerryPRA2009, WiebePRL2016}, 
including those using GHZ states~\cite{BerryPRA2009, KesslerPRL2014, KomarNATPHY2014, PezzeEPL2007, MitchelSPIE2005}.
Moreover, our protocol is implemented with a {\it single} measurement of a collective spin observable at each step
and the phase is extracted with a practical estimation technique. 
 
 \section{Results}
 
 In the following, we present our QPE algorithm and discuss its performance in the ideal case and 
in presence of decoherence.
Mathematical details are reported in the Methods and in the Supplementary Information (SI). \\

{\bf Gaussian spin states quantum phase estimation algorithm.}
The circuit representation of the $K+1$ steps of the algorithm is shown in Fig.~\ref{fig1}(a), while Fig.~\ref{fig1}(b) illustrates a single step.
The phase estimation uses $N_T$ particles that can access two internal or spatial modes. 
The ensemble of $N_T$ qubits is divided into $K+1$ spin-polarized states 
$\ket{0}^{\otimes N_k}$, one for each step $k=0,1,...,K$ of the algorithm, with $N_k \gg 1$ and $\sum_{k=0}^K N_k = N_T$.
We also introduce collective spin operators $\J_{n} =\sum_{j=1}^{N_k} \hat{\sigma}_{n}^{(j)}/2$,
where $\hat{\sigma}_{n}^{(j)}$ is the Pauli matrix of particle $j$ along the axis $n=x,y,z$ in the Bloch sphere.
Each state $\ket{0}^{\otimes N_k}$ first goes through a collective Hadamard gate $H^{\otimes N_k}$, 
which prepares the coherent spin state $\big( \tfrac{\ket{0}+\ket{1}}{\sqrt{2}} \big)^{\otimes N_k}$.
Except at the zeroth step $k=0$, the state then goes through a spin-squeezing gate that generates the GSS
$\ket{\psi_k}$ with squeezing parameter $s_k^2 = 4 (\Delta \J_y)^2_k/N_k$ (see Methods).
Notice that the spin-squeezing gate creates entanglement among the $N_k$ ancilla qubits~\cite{SorensenNATURE2001, PezzePRL2009}.
This concludes the state preparation. 
It should be noticed that the spin-squeezing gate (represented in Fig.~\ref{fig1} by the green box), can implemented 
experimentally in a variety of ways and experimental systems, see Ref.~\cite{PezzeRMP2018} for a review.

The phase encoding is obtained from a controlled-$U$ gate applied to each qubit. 
The gate gives a phase shift $\theta$ to the qubit in the state $\ket{1}$, while leaving $\ket{0}$ unchanged~\cite{NielsenBOOK, ClevePRSA1998, AbramsPRL1999}:
more explicitly, $U_c \ket{1} \ket{u} = e^{-i \theta} \ket{1} \ket{u}$ and $U_c \ket{0} \ket{u} = e^{-i \theta} \ket{0} \ket{u}$, where
$\ket{u}$ is an eigenstate of $U$ which is stored in the register and $e^{-i \theta}$ is the corresponding eigenvalue.
Let us write the spin-squeezed state as $\ket{\psi_k} = \sum_{\mu_k=-N_k/2}^{N_k/2} c_k(\mu_k) \ket{\mu_k}_z$, 
where $\ket{\mu}_z$ are eigenstates of $\J_z$ with eigenvalues $-N_k/2 \leq \mu_k \leq N_k/2$, 
given by the symmetrized combination $\ket{\mu}_z = {\rm Sym}[\ket{0}^{\otimes (N_k/2+\mu_k)} \otimes \ket{1}^{\otimes (N_k/2-\mu_k)}]$, 
and $c_k(\mu_k)$ are Gaussian amplitudes (see Methods).
Overall, the controlled-$U^{\otimes N_k}$ operation applied to $\ket{\psi_k}$ gives 
$U_c^{\otimes N_k} \ket{\psi_k} \ket{u} = \sum_{\mu_k=-N_k/2}^{N_k/2} c_k(\mu_k) e^{-i (N_k/2-\mu_k) \theta} \ket{\mu_k}_z \ket{u}$ and is equivalent to a 
collective spin rotation of the state $\ket{\psi_k}$ by the angle $\theta$ around the $z$ axis, namely $U_c^{\otimes N_k} \ket{\psi_k} \ket{u} = e^{-i \theta \J_z} \ket{\psi_k} \ket{u}$.
The algorithm requires, in total, $N_T$ controlled-$U$ gates.

The readout consists, first, of a collective $\pi/2$ rotation around the $x$ axis, 
 $R_x^{\otimes N_k} = e^{-i  \tfrac{\pi}{2} \J_x}$, and a final projection along the $z$ axis (namely the measurement of $\J_z$) 
 with possible result $-N_k/2 \leq \mu_k \leq N_k/2$.
The single measurement leads to the estimate $\theta_{\rm est}^{(k)}(\mu_k) = \arcsin (\mu_k/\langle \psi_k \vert \J_x \vert \psi_k \rangle)$, see Methods.

The key operation of the algorithm is the phase feedbacks (represented by pink lines and boxes in Fig.~\ref{fig1}): 
before readout, the state $e^{-i \theta \J_z} \ket{\psi_k}$ is sequentially rotated by
$e^{i  \J_z \theta_{\rm est}^{(0)}}~e^{i  \J_z \theta_{\rm est}^{(1)}}...e^{i \J_z \theta_{\rm est}^{(k-1)}} \equiv \bigotimes_{j=0}^{k-1} R_z(-\theta_{\rm est}^{(j)})$,
with $\theta_{\rm est}^{(j)}$ being the estimated value at the step $j<k$.
This rotation places the GSS $\ket{\psi_k}$ close to its optimal sensitivity point, namely on the equator of the generalized Block sphere, see Fig.~\ref{fig1}(b).
After $K$ steps the circuit outputs are $K+1$ values $\theta_{\rm est}^{(0)}, ... \theta_{\rm est}^{(K)}$:
their sum, $\theta_{{\rm est},K} = \sum_{k=0}^{K} \theta_{\rm est}^{(k)}$, estimates the unknown $\theta$.
Notice that the circuit described above leads to an estimate of $\theta$ within the inversion region $[-\pi/2, \pi/2]$ of the $\arcsin$ function.
The algorithm can be extended to the full range $\theta \in [-\pi, \pi)$ by a slight modification of only the zeroth step, see Methods and SI. 
Below, we analyze the sensitivity of the QPE for an arbitrary $\theta\in [-\pi, \pi)$.\\

{\bf Phase sensitivity.}
The sensitivity in the estimation of $\theta$, $\Delta^2 \theta_{{\rm est},K}$, is
quantified by the statistical variance of $\theta_{{\rm est},K} -\theta$.
As shown in the Methods, this is calculated using the recursive formula
\be \label{eq.Dtheta}
\Delta^2 \theta_{{\rm est},k} = 
\frac{s^2_k}{N_k} + \frac{\Delta^2 \theta_{{\rm est},k-1}}{2 s_k^4 N_k^2},
\ee
for $k=0, 1, ...K$ and initial condition $\Delta^2 \theta_{{\rm est},0} = 4/N_0$.
Equation~(\ref{eq.Dtheta}) assumes $s_k^2 N_k \gg 1$. 
While this is only partially justified for large values of $k$, it is nevertheless in excellent agreement 
with full numerical results. Higher order terms are explicitly calculated in the SI.
The optimization of Eq.~(\ref{eq.Dtheta}) over $s_1, ..., s_K$ and $N_0, ..., N_K$ can be performed analytically (with $s_0=1$ and fixing the total number of qubits $N_T$)
leading to
\be \label{eq.Nj}
N_{k} = 4 \times 3^{k-1} N_0,
\ee
and
\be \label{sopt}
s_k^2 = 
\frac{3^{\tfrac{5}{2}-\tfrac{3}{2}\tfrac{1}{3^{k-1}}-k}}{2^{\tfrac{7}{2}-\tfrac{5}{2}\tfrac{1}{3^{k-1}}}} \frac{1}{N_0^{1-1/3^k}}.
\ee
In particular, the first step of the protocol is implemented with a GSS containing $N_1 = 4 N_0$ particles and,  
at each further step, the number of particles is increased by a factor 3: $N_{k+1} = 3 N_{k}$, for $k=0, ..., K-1$. 
Using Eq.~(\ref{eq.Nj}) and summing the geometric series $N_T = \sum_{k=0}^{K}N_k$, we obtain $N_0 = N_T/(2 \times 3^K - 1)$.
The protocol uses GSS that are more and more squeezed (namely, with $s_k$ decreasing with $k$) as the number of steps increases, while $s^2_k N_k$ 
saturates to the asymptotic value $\sqrt{27/2}$ for $k\gg 1$, see SI.
The analytically-optimized sensitivity is 
\be \label{eq.Dthetafinal}
\Delta \theta_{{\rm est},K}= \frac{\alpha_K}{ N_T^{1-1 /(2\times 3^K)}}
\ee
that very rapidly (in the number of steps $K$) approaches the Heisenberg scaling with respect to
the total number of qubits and with a prefactor that converges to $\alpha_{K \to \infty} = 4$.
It is also worth noticing that the sensitivity is exponential in the number of steps $K \sim \log_3 N_T$, 
while with standard QPE protocols $K \sim \log_2 N_T$~\cite{Kitaev, KitaevBOOK, HigginsNATURE2007, BerryPRA2009}.
Already for $K=3$ (that uses one coherent spin state and three spin-squeezed states with decreasing $s_k$) we obtain a sensitivity 
$O(N_T^{-0.98})$, which is very close to the Heisenberg limit.

%%%%%%%%%%%%%%%%%%%%%%%%%%%%%%%%%%%%%%%%%%%%%%%%%%%%%%%%%%%%%%%%%%%%%%%%%%%%%%
%% Figure 2
%%%%%%%%%%%%%%%%%%%%%%%%%%%%%%%%%%%%%%%%%%%%%%%%%%%%%%%%%%%%%%%%%%%%%%%%%%%%%%
\begin{figure} [t!]
\centering
\includegraphics[width=1\columnwidth]{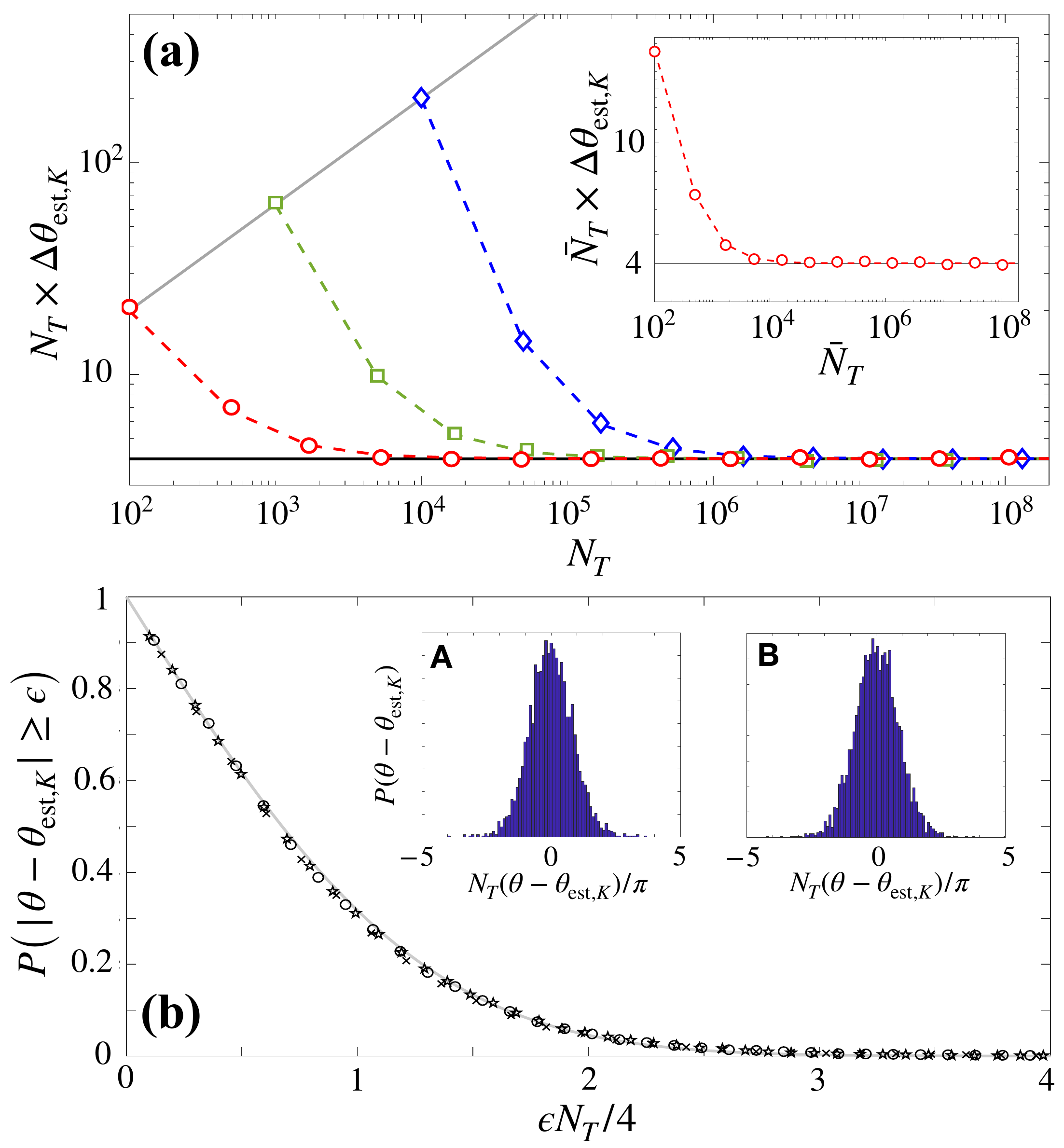}
\caption{
{\bf Phase sensitivity of the QPE.}
(a) Phase sensitivity $\Delta \theta_{{\rm est},K}$ as a function of the total number of particles $N_T$ and for initial $N_0=10^2$
(red lines and symbols), $N_0=10^3$ (green) and $N_0=10^4$ (blue). 
Symbols are results of numerical Monte-Carlo simulations where $\theta$ is chosen randomly in $[-\pi, \pi)$.
The dotted lines are guide to the eye connecting analytical predictions, see text and SI.
The grey line is $\Delta \theta_{\rm est} = 2/\sqrt{N_0}$ while the horizontal solid line is $\Delta \theta_{\rm est} = 4/N_T$.
The inset show the results of simulations obtained with
states containing a fluctuating number of particles (see text), where $\bar{N}_T$ is the total average number of particles
and $\bar{N}_0=100$. 
(b) Error probability $P(\vert \theta -  \theta_{{\rm est},K}   \vert \geq \epsilon)$ as a function of $\epsilon N_T /4$.
Symbols are Monte-Carlo results obtained with $N_0 =10^2$ and $K=8$ (crosses), $K=10$ (stars), $K=13$ (circles). 
The solid grey line is $1-  {\rm erf} [\epsilon N_T/(4\sqrt{2})]$.
The insets show the distributions $P(\theta - \theta_{{\rm est},K})$ for $N_0=100$, $K=8$ (A) and $K=13$ (B).
Results in all panels are obtained from $10^4$ independent repetition of the protocol.}
\label{fig2}
\end{figure} 
%%%%%%%%%%%%%%%%%%%%%%%%%%%%%%%%%%%%%%%%%%%%%%%%%%%%%%%%%%%%%%%%%%%%%%%%%%%%%%
%%%%%%%%%%%%%%%%%%%%%%%%%%%%%%%%%%%%%%%%%%%%%%%%%%%%%%%%%%%%%%%%%%%%%%%%%%%%%%
%%%%%%%%%%%%%%%%%%%%%%%%%%%%%%%%%%%%%%%%%%%%%%%%%%%%%%%%%%%%%%%%%%%%%%%%%%%%%%

In Fig.~\ref{fig2}(a) we show the results of numerical Monte-Carlo simulations of our QPE algorithm where $\theta$ is chosen randomly in $[-\pi, \pi)$ with a flat distribution.
For further clarity, we report the QPE pseudo-code in the Methods section.
The sensitivity $\Delta \theta_{{\rm est},K}$ approaches the Heisenberg scaling Eq.~(\ref{eq.HL})  
 as a function of the number of qubits (or, equivalently, the number of steps $K+1$), 
independently of the initial $N_0$.
This is confirmed by $N_T \times \Delta \theta_{{\rm est},K}$ approaching the constant value 4 for large $N_T$ and different initial $N_0$ 
[colored symbols in Fig.~\ref{fig2}(a)].
Numerical simulations agree well with the analytical optimization (colored dashed lines) results.

In Fig.~\ref{fig2}(b) we analyze the behaviour of the estimator, $\theta_{{\rm est},K}$.
As shown by the insets of Fig.~\ref{fig2}(b), the distribution of $\theta - \theta_{{\rm est},K}$ is -- to a very good approximation --
a Gaussian cantered in $0$, namely the estimator is statistically unbiased (notice that the histograms are obtained for $10^4$ independent repetition of the algorithm for random $\theta$).
Furthermore, the probability of making an error $\epsilon$ in the estimation of an arbitrary $\theta$ is 
\be
P(|\theta_{{\rm est},K} - \theta \vert \geq \epsilon) = 1- {\rm erf}[\epsilon N_T /(4\sqrt{2})],
\ee
where ${\rm erf}(x)$ is the error function, see Fig.~\ref{fig2}(b).
The error probability is thus exponentially small with $N_T$. \\

\begin{figure} [t!]
\centering
\includegraphics[width=1\columnwidth]{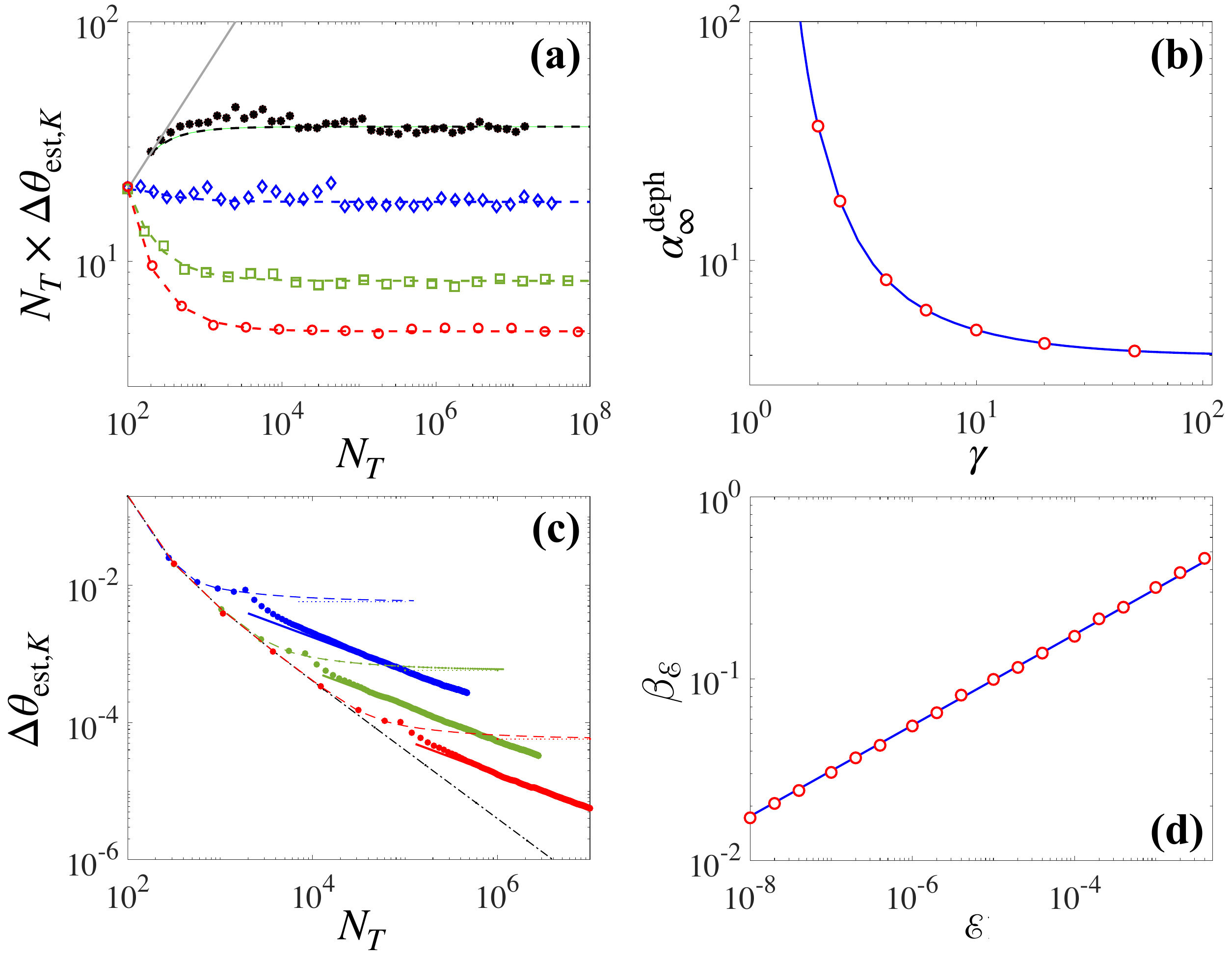}
\caption{{\bf Sensitivity of the QPE in presence of noise.}
Panels (a) and (b) consider the effect of collective dephasing described by Eq.~(\ref{dephasing}) with 
$P(\phi) \approx e^{\gamma \cos \phi}$.
Panel (a) reports $\Delta \theta_{{\rm est},K}$ as a function of the total number of particles $N_T$ and for different values of $\gamma$:
$\gamma = 10$ (red lines and symbols), $4$ (green), $2.5$ (blue) and $2$ (black). 
Symbols are results of Monte-Carlo simulations, while dotted lines are expected results (see SI).
Panel~(b) shows the prefactor $\alpha_{\infty}^{\rm deph}$ of the Heisenberg scaling in Eq.~(\ref{HeisenbergDep}) as a function of $\gamma$.
Circles are results of Monte-Carlo simulations (obtained asymptotically in the number of steps), the solid line is the expected theoretical result, see SI.
In panels~(c) and~(d) we consider full depolarization.
Panel~(c) shows $\Delta \theta_{{\rm est},K}$ as a function of $N_T$: symbols are results of simulations for different values of 
$\mathcal{E}$: $\mathcal{E} = 10^{-4}$ (blue circles), $10^{-6}$ (green) and $10^{-8}$~(red).
Lines are the analytical results (see text). The black dashed line is the noiseless case.
Asymptotically in $N_T$, the sensitivity follows Eq.~(\ref{HeisenbergDepol}) with prefactor $\beta_{\mathcal{E}}$ shown 
in panel (d) as a function of $\mathcal{E}$.
Here the blue line is $\beta_{\mathcal{E}} = c \, \mathcal{E}^{1/4}$ where $c =1.75$ is obtained from a fit.
Red circles are results of numerical simulations.
}
\label{fig3}
\end{figure}

{\bf Impact of decoherence.}
We now investigate the robustness of the protocol against noise and decoherence in realistic experimental implementations.
We include a noise source in the GSS and assume ideal phase rotations (which are typically implemented on time scales much shorter than squeezed-state preparations).
We first consider a collective dephasing along an arbitrary axis $n$ in the Bloch sphere, which is described by the transformation
\be \label{dephasing}
\Lambda_{n}\big[\ket{\psi_k}\big] = \int_{-\pi}^{\pi} d\phi ~ P(\phi) ~ e^{- i \phi \J_n} \ket{\psi_k} \bra{\psi_k}  e^{i \phi \J_n}.
\ee
This provides a stochastic rotation $e^{- i \phi \J_n}$ with an angle $\phi$ distributed with probability $P(\phi)$,
where $\ket{\psi_k}$ is the ideal GSS.
Without loss of generality (upon a further proper rotation of the state $\ket{\psi_k}$) we consider depolarization along the $n=y$ axis.
The noise leaves unchanged the moments of $\J_y$ -- in particular it does not affects the squeezing along the $y$ axis -- 
but it decreases the length of the collective spin, $\mean{\J_x}$, while increasing the bending of the state in the sphere, namely $\mean{\J_x^2}$
(see SI for details).
Results of numerical simulation of our QPE algorithm are shown in Fig.~\ref{fig3}(a) and (b). 
For a sufficiently large number of steps, the protocol reaches the Heisenberg scaling (for $K \gg 1$)
\be \label{HeisenbergDep}
\Delta \theta_{\rm est} = \frac{\alpha_{\infty}^{\rm deph}}{N_T}
\ee
where the prefactor $\alpha_{\infty}^{\rm deph}$ is determined by the low lying Fourier components 
$\int_{-\pi}^{\pi} d\phi \cos (2^\lambda \phi) P(\phi)$, with $\lambda=0,1$, see SI. 
Notice that the Heisenberg scaling (\ref{HeisenbergDep}) is recovered even when the width of $P(\phi)$ is of the order of 
$2\pi$, highlighting the robustness of the QPE algorithm to this source of noise.
This is in contrast with QPE protocols implemented with GHZ states, where there is no
preferred rotation axis that guarantees robustness to dephasing noise.  

%%%%%%%%%%%%%%%%%%%%%%%%%%%%%%%%%%%%%%%%%%%%%%%%%%%%%%%%%%%%%%%%%%%%%%%%%%%%%%
%% Figure 3
%%%%%%%%%%%%%%%%%%%%%%%%%%%%%%%%%%%%%%%%%%%%%%%%%%%%%%%%%%%%%%%%%%%%%%%%%%%%%%
%%%%%%%%%%%%%%%%%%%%%%%%%%%%%%%%%%%%%%%%%%%%%%%%%%%%%%%%%%%%%%%%%%%%%%%%%%%%%%
%%%%%%%%%%%%%%%%%%%%%%%%%%%%%%%%%%%%%%%%%%%%%%%%%%%%%%%%%%%%%%%%%%%%%%%%%%%%%%
%%%%%%%%%%%%%%%%%%%%%%%%%%%%%%%%%%%%%%%%%%%%%%%%%%%%%%%%%%%%%%%%%%%%%%%%%%%%%%

The situation is different in presence of full depolarization (within the symmetric subspace of dimension $N+1$), described by the transformation
\be
\Lambda_{\mathcal{E}}\big[\ket{\psi_k}\big] =(1-\mathcal{E})\ket{\psi_k} \bra{\psi_k}  + \frac{\mathcal{E}}{N+1}  \Eins,
\ee
where $0 \leq \mathcal{E}\leq 1$ is the depolarization parameter.
This noise affects the spin moments along all directions and, in particular, it poses a limit
to the smallest achievable squeezing parameter $s_{\rm min}^2 = \mathcal{E} N/3$, see SI.
Simulation of the QPE in presence of full depolarization are shown in Fig.~\ref{fig3}(c) and (d).
Similar results holds for different noise models which pose limitations to the maximum available squeezing, e.g. particles losses (see SI).  
The QPE protocol in presence of noise follows the ideal (noiseless) phase estimation sensitivity  
up to a number of steps $k \lesssim \tilde{k}$ for which $s_k \geq s_{\rm min}$.
In particular, if $\mathcal{E}  \ll 1$ such that $\tilde{k} \gg 1$, we recover the Heisenberg scaling~(\ref{eq.HL}) in a finite range of total number of particles.
The fact that depolarization noise is irrelevant up to a critical value of squeezing $s_{\rm {min}}$ and number of particles $N_{\tilde k}\sim 4 \times 3^{\tilde k} N_0$
is the basic physical reason behind the expected experimental robustness of our phase estimation algorithm.
When $k \geq \tilde{k}$, it is more efficient to avoid phase feedback between different steps 
[which would lead to a saturation of $\Delta \theta_{\rm est} \times N_T$ to the asymptotic value $\sqrt{\mathcal{E}/3}$, see dashed lines in Fig.~\ref{fig3}(c)]
and rather repeat the protocol with multiple copies of 
the squeezed state $\ket{\psi_{\tilde{k}}}$ of $N_{\tilde{k}}$ particles and squeezing parameter $s_{\tilde{k}}$.
Asymptotically in $N_T$ we thus reaches a sensitivity
\be \label{HeisenbergDepol}
\Delta \theta_{\rm est} = \frac{\beta_{\mathcal{E}}}{\sqrt{N_T}},
\ee
with a prefactor $\beta_{\mathcal{E}} \sim \mathcal{E}^{1/4}$ (see SI).
Symbols in Fig.~\ref{fig3}(c) show results of numerical Mote-Carlo simulations, in excellent agreement with analytical calculations.
In panel (d) we show $\beta_{\mathcal{E}}$ as a function of $\mathcal{E}$ where the results of simulations (circles) are compared to 
the expected scaling with $\mathcal{E}$. \\

We have so far illustrated the protocol with quantum states having fixed and known total number of particles. 
In some experimental implementations, for instance with ultracold atoms, it is possible to fix the number of particles 
only in average, provided several identical preparations of the sample, with a fluctuating, unknown, number of atoms in a single
realisation. Our protocol is unaffected by 
these fluctuations: in the inset of \ref{fig2}(a) we plot $\Delta \theta_{{\rm est},K}$ as a function of the total average number of particles $\bar{N}_T$, 
 where the protocol at the $k$th step is implemented with states having shot-noise particle-number fluctuations $(\Delta \hat{N}_k)^2  = N_k$.
We also emphasize that using the non-linear readout~\cite{DavisPRL2016, FrowisPRL2016, MacriPRA2016} (which has been demonstrated experimentally for atomic squeezed states~\cite{HostenSCIENCE2016}), our QPE protocol can be made robust against detection noise. 
To conclude, we notice that in some applications as in atomic clocks~\cite{KesslerPRL2014, KomarNATPHY2014, LudlowRMP2015}, 
it is necessary to maximize the sensitivity by implementing each round of the protocol with 
the maximum possible number of particles available, namely $N_k = N$ for all $k$~\cite{PezzePRL2020}.
%The corresponding optimization of the squeezing parameters is reported in Ref.~\cite{PezzePRL2020}.

\section{Discussion}
 
To summarize, we have proposed a novel QPE algorithm that uses Gaussian spin states and 
reaches a sensitivity at the Heisenberg limit $O(1/N_T)$, with respect to the total number of qubits $N_T$ or, equivalently, the total number of applications of a controlled-$U$ gate.
There are two main differences with respect to the standard QPE algorithms~\cite{Kitaev, KitaevBOOK} using single ancilla qubits~\cite{HigginsNATURE2007, WiebePRL2016}, 
including those based on the inverse quantum Fourier transform~\cite{ClevePRSA1998, NielsenBOOK, Aspuru-GuzikSCIENCE2005}: 
i) The number of controlled-$U$ gates in the GSS algorithm of Fig.~\ref{fig1} scales linearly with $N_T$ rather than exponentially. 
This means that, if we take into account the time $t_U$ necessary to implement a single controlled-$U$ gate, the GSS algorithm is exponentially faster.
This property is also shared by QPE algorithms using GHZ states~\cite{PezzeEPL2007, BerryPRA2009}, that are however more fragile to noise.
The short time required by the algorithm makes it well suited for the estimation of time-varying signals. 
More explicitly, if we indicate with $\alpha = d\theta(t)/dt$ the local slope of the signal, 
we obtain  the condition $\alpha \lesssim O(t_U N_T^2)^{-1}$ for the Heisenberg limited estimation  of the time-varying $\theta$ with the GSS algorithm, 
compared to more stringent condition $\alpha \lesssim O(t_U N_T e^{N_T})^{-1}$ for the QPE algorithms using single ancilla qubits.
ii) Differently from Kitaev's QPE protocol~\cite{Kitaev, KitaevBOOK}, the phase estimation at each step of the GSS algorithm is based on a single measurement. 
The knowledge about $\theta$ is progressively sharpened using GSS of higher number of particles and decreasing squeezing parameter.
The key operation of the algorithm is the phase feedback that can be understood as an adaptive measurement~\cite{WisemanBOOK} able to place the 
GSS around its most sensitive point.
In particular, the analytical optimization provided by Eqs.~(\ref{eq.Nj}) and (\ref{sopt}) allow to fully pre-determine the number of particle and the strength of the squeezing gate at each step.
Therefore, the adaptive measurement in the GSS algorithm does not require the numerical optimization of states,
operations and/or control phases~\cite{BerryPRA2009, WiebePRL2016}, nor the support of any classical memory to store the phase distribution~\cite{HigginsNATURE2007}.

Commonly to all QPE algorithms, our protocol uses controlled-$U$ gates to map a quantity of interest to a phase to be estimated. 
The GSS algorithm thus shares all known applications of QPE~\cite{ShotSIAM1997, AbramsPRL1999, McArdleRMP2020}, while using noise-resilient quantum states.
The use of GSS, which are routinely created in labs, can open a novel route for experiments with cold and ultracold atoms toward 
applications in quantum computing, quantum computational chemistry and quantum simulation. 
 
\section{Methods}

{\bf Estimation method.}
The phase estimation protocol of the QPE algorithm follows the standard method of moment~\cite{PezzeRMP2018}.
The output state is characterized by an average collective spin moment $\mean{\J_z^{\rm out}}$
that can be expressed as a function of the mean spin value $\mean{\J_x}$ of the state at the end of the state-preparation step 
(i.e. before phase imprinting),
$\mean{\J_z^{\rm out}} = \mean{\J_x} \sin \theta$.
From a single measurement of $\J_z^{\rm out}$ with result $\mu$, we estimate $\theta$ as
\be \label{estimator}
\theta_{\rm est}(\mu) = \arcsin \big( \mu / \mean{\J_x} \big).
\ee 
The sensitivity of this estimator is given by the statistical variance
$\Delta^2 \theta_{\rm est} = \sum_{\mu=-N/2}^{N/2} P(\mu\vert \theta) \big( \theta_{\rm est}(\mu) - \bar{\theta}_{\rm est} \big)^2$, 
where $\bar{\theta}_{\rm est} = \sum_{\mu=-N/2}^{N/2} P(\mu\vert \theta) \theta_{\rm est}(\mu)$ is the statistical mean value
and $P(\mu\vert \theta)$ is the probability to obtain the result $\mu$ for a given $\theta$.
The statistical variance can be well approximated by the prediction of error propagation (see SI)
\be \label{variance}
\Delta^2 \theta_{\rm est} \approx \frac{(\Delta \J_z^{\rm out})^2}{(d \mean{\J_z^{\rm out}}/d \theta)^2} =
\frac{(\Delta \J_y)^2}{\mean{\J_x}^2} +  \frac{(\Delta \J_x)^2}{\mean{\J_x}^2} \tan^2 \theta,
\ee
where we have taken into account that $\mean{\J_z}=0$ and $\mean{\J_z \J_x}=0$ for the GSS
considered in the manuscript. \\

{\bf Analytical calculation of the spin moments.}
We assume that the spin-squeezing gate in Fig.~\ref{fig1} generates the Gaussian state 
\be \label{Gaussian}
\ket{\psi(N,s)} = \frac{1}{\sqrt{\mathcal{N}}} \sum_{\mu=-N/2}^{N/2} e^{-\tfrac{\mu^2}{s^2N}} \ket{\mu}_y,
\ee
starting from a coherent spin state 
$(\ket{0} + \ket{1})^{\otimes N}/2^{N/2}$ of $N$ qubits.
One-axis-twisting~\cite{KitagawaPRA1993} and non-destructive measurements (e.g. by atom-light interaction~\cite{HammererRMP2010}) generate spin-squeezed states that can be well approximated by Eq.~(\ref{Gaussian}).
Here, $s$ is a squeezing parameter ($s<1$ for spin-squeezed states), $\ket{\mu}_y$ is the eigenstates of $\J_y$ with eigenvalue $\mu$ ($\mu = -N/2, -N/2+1, ..., N/2$)
and $\mathcal{N}$ is provides the normalization.
As detailed in the SI, we can calculate analytically, to a very good approximation, mean values and variances of $\J_{x,y,z}$ for the state (\ref{Gaussian}).
For $s^2N \gtrsim 1$, we find 
$\mean{\J_x} \approx N e^{-1/(2s^2 N)}/2$,  $\mean{\J_y} = \mean{\J_z} = 0$,
$\mean{\J_x^2} \approx N^2 ( 1+ e^{-2/(s^2 N)})/8$,  
$\mean{\J_y^2} = Ns^2/4$ and 
$\mean{\J_z^2} \approx N^2 ( 1- e^{-2/(s^2 N)})/8$. 
A comparison between the analytical results and exact numerical calculations is reported in the SI.
Notice that, for $s=1$, we recover the well known spin moments of a coherent spin state $(\ket{0} + \ket{1})^{\otimes N}/2^{N/2}$, 
namely $\mean{\J_x} = N /2$, $\mean{\J_x^2} = \mean{\J_y^2} = \mean{\J_z^2} = N^2/4$, a part corrections of the order $1/N$. 
Replacing the collective spin moments into Eq.~(\ref{variance}) we obtain
\be \label{DthetaMet}
(\Delta \theta)^2 = \frac{2s^2 + N ( 1 - e^{-1/(s^2 N)} )^2 \tan^2 \theta}{2N e^{-1/(s^2 N)}}.
\ee
This equation can be generalized in presence of noise by calculating the spin moments for the noisy state and replacing them into Eq.~(\ref{variance}), see SI. \\

{\bf Phase estimation protocol.}
We now discuss the different steps of the phase estimation protocol. Further mathematical details are reported in the SI. \\

{\it Zeroth step.} It is implemented with a coherent spin state of $N_0$ particles without requiring a spin-squeezing gate.
The state is rotated by $e^{-i (\theta/2) \J_z}$ where the factor 2 dividing the rotation angle $\theta$ in the phase encoding transformation guarantees that
the behaviour of $\mean{\J_z^{\rm out}}$ is monotonic as a function of $\theta$ in the full phase interval $[-\pi, \pi)$.
The estimation method discussed above provides the estimator $\theta_{\rm est}^{(0)}(\mu_0) = 2 \arcsin ( 2\mu_0 / N_0)$, depending on the measurement results $\mu_0$,
with a sensitivity $\Delta^2 \theta_{{\rm est},0}  = 4/N_0$ (notice that $(\Delta \J_x)^2=0$ for the coherent spin stata). 
The factor $4$ in the sensitivity above the standard quantum limit is a direct consequence of the factor 2 dividing the rotation angle.\\

{\it $k$th step.} The state preparation of the $k$th step ($k=1, ..., K$) provides the spin-squeezed state $\ket{\psi_k}$ of $N_k$ particles and squeezing parameter $s_k$, see Eq.~(\ref{Gaussian}).
The state is transformed by the controlled-$U^{\otimes N_k}$ gate and a series of $k$ rotation gates 
$\prod_{j=0}^{k-1} R_z(-\theta_{\rm est}^{(j)})$ gate (which uses the estimated value $\theta_{\rm est}^{(j)}$ obtained in the previous steps of the protocol, $j<k$).
The overall rotation applied to the spin-squeezed state is $e^{-i \theta_k \J_z}$, where $\theta_k = \theta - \sum_{j=0}^k \theta_{\rm est}^{(j)}$ is a stochastic variable with distribution 
$P(\theta_k) \sim e^{-\theta_1^2/(2\kappa_{k-1}^2)}$, with $\kappa_{k-1}^2 = \Delta^2 \theta_{{\rm est},k-1}$.
A measurement of $\J_z$ after a final rotation $R_x(\pi/2)$ provides a result $\mu_k$ and a corresponding estimate $\theta_{\rm est}^{(k)}(\mu_k)$, according to Eq.~(\ref{estimator}).
The value $\theta_{\rm est}^{(k)}(\mu_k)$ is used to implement the adaptive phase rotation at the ($k+1$)th step. \\

{\it Phase sensitivity.}
Assuming that the protocol is stopped after $K$ steps, 
The phase $\theta$ is estimated by $\theta_{{\rm est},K}= \sum_{k=0}^K \theta_{\rm est}^{(k)}$ 
(notice that $\theta_{\rm est}^{(K)}-\theta_K = \theta_{{\rm est},K} - \theta$, where $\theta_K = \theta - \sum_{j=0}^K \theta_{\rm est}^{(j)}$ is the overall
phase rotation at the $K$th step).
The corresponding sensitivity is obtained by a statistical average obtained by integrating Eq.~(\ref{DthetaMet}) (with the replacements $s, N, \theta \to s_K, N_K, \theta_K$, and approximating $\tan^2 \theta_K \approx \theta_K^2$) over the distribution $P(\theta_K)$ of $\theta_K$.
We obtain
\be \label{EDthetan}
 \Delta^2 \theta_{{\rm est},K}  =
 \frac{2s_K^2 + N_K ( 1 - e^{-1/(s_K^2 N_K)} )^2  \Delta^2 \theta_{{\rm est},K-1}  }{2N_K e^{-1/(s_K^2 N_K)}}
\ee 
giving Eq.~(\ref{eq.Dtheta}), to the leading order in $s_k^2 N_k \gg 1$.
The recursive relation, with initial condition $\Delta^2 \theta_{{\rm est},0} = 4/N_0$
provides a set of $\Delta^2 \theta_{{\rm est},k}$ with $k=1, ..., K$.
The optimization of Eq.~(\ref{EDthetan}) over $s_1, s_2, ..., s_K$ and $N_0, N_1, ..., N_K$ is as follows.
First, we minimize Eq.~(\ref{EDthetan}) with respect to $s_K$, that gives 
\be
s_K^2 N_K = ( N_K^2 \Delta^2 \theta_{{\rm est},K-1} )^{1/3},
\ee
This equation is also understood as a recursive relation giving the squeezing parameters $s_1, s_2, ..., s_K$ as a function of $N_0, N_1, ..., N_K$, with initial condition $s_0=1$
(for the coherent spin state).
Using the optimal values $s_1, ..., s_K$, Eq.~(\ref{eq.Dtheta}) becomes
\be
 \Delta^2 \theta_{{\rm est},K} =  \bigg( \frac{3}{2} \bigg)^{\tfrac{3}{2}\big(1-\tfrac{1}{3^K}\big)}
\frac{4^{1/3^K}}{N_{K}^{4/3} N_{K-1}^{4/9} ... N_{1}^{4/3^K} N_0^{1/3^K}},
\ee 
which can be further optimized as a function of $N_0, N_1, ..., N_K$ with the constraint of a fixed $N_T = \sum_{k=0}^K N_k$.
We obtain a set of $K$ linear equations which can be recast in the matrix form
$( \vect{A} + \vect{u} \vect{u}^T) \cdot \vect{x} = N_T \vect{u}$,
where
$\vect{x} = (N_K, ..., N_1)$, 
$\vect{u} = (1, 1, ..., 1)$,
and 
\be
\vect{A} = 
\frac{1}{4}
\begin{pmatrix}
\frac{1}{3^{K-1}} & 0 & 0 &... & 0\\
0 & \frac{1}{3^{K-2}} & 0 & ... &0 \\
0 & 0 & \frac{1}{3^{K-3}}  & ... & 0 \\
\vdots & \vdots & \vdots & \ddots & \vdots \\ 
0 & 0 & 0 & ... & 1
\end{pmatrix}.  \nonumber
\ee
The solution of the linear set of equation is found using the Sherman-Morrison formula, giving 
$\vect{x} = \frac{N_T}{2 \times 3^K - 1} \vect{A}^{-1}\vect{u}$, and leading to 
Eqs.~(\ref{eq.Nj}) and (\ref{sopt}).\\

{\bf Pseudo-code of the algorithm.} 
For further clarity, we report below the pseudo-code of the algorithm. \\

\noindent\rule{6cm}{0.4pt} \\
%\noindent\rule{9cm}{0.8pt} 
$\theta =$ \text{generate\textunderscore random} $\in[-\pi, \pi)$ \\
%\noindent\rule{7cm}{0.4pt} \\
\noindent {\bf input}: $N_0$ or $N_T$, $K$\\
{\bf for } $k=0,1\dots,K$: \\
\hspace*{.8cm} {\bf if} $k=0$ \\
\hspace*{1.6cm} $\theta_k = \theta/2$ \\
\hspace*{.8cm} {\bf elseif} $k \neq 0$ \\
\hspace*{1.6cm} $\theta_k = \theta - \sum_{j=0}^{k-1} \theta_{\rm est}^{(j)}$ \\
\hspace*{.8cm} {\bf end} \\
\hspace*{.8cm} $\mu_k$ = \text{generate\textunderscore random} \text{from} $P(\mu_k \vert \theta_k)$ \\
\hspace*{.8cm} {\bf if} $k=0$ \\
\hspace*{1.6cm} $\theta_{\rm est}^{(0)} = 2\arcsin \big( 2\mu / N_0 \big)$ \\
\hspace*{.8cm} {\bf elseif} $k \neq 0$ \\
\hspace*{1.6cm} $\theta_{\rm est}^{(k)} = \arcsin \big( 2\mu / \bra{\psi_k} \J_z \ket{\psi_k}\big)$ \\
\hspace*{.8cm} {\bf end} \\
\hspace*{.8cm} \text{update} $N_k$ [Eq.~(\ref{eq.Nj})], $s_k$ [Eq.~(\ref{sopt})]\\
{\bf end}\\
%\noindent\rule{7cm}{0.4pt} \\
{\bf return}: $\theta_{\rm est,K} = \sum_{k=0}^{K} \theta_{\rm est}^{(k)}$\\
\noindent\rule{6cm}{0.4pt} \\

Notice that, as initial conditions, we can either fix the total number of qubits $N_T$ or the number of qubits in zeroth-step state, $N_0$.
In the former case, the code starts with $N_0 = N_T/(2 \times 3^K - 1)$, in the latter case, it uses total $N_T =  (2 \times 3^K - 1) N_0$ qubits.
Given $N_T$ qubits, one can further optimize the algorithm over the number of steps $K$, as shown in the SI.

Numerical Monte-Carlo simulations of the QPE algorithm require the probability distribution $P(\mu_k \vert \theta_k)$.
%while for the experimental implementation, there is no need to access such distribution.
It can be calculated exactly as $P(\mu_k \vert \theta_k) = \big\vert {}_z\bra{\mu} e^{-i (\pi/2) \J_x} e^{-i \theta_k \J_z} \ket{\psi_k} \big\vert^2$, 
up to $N_k \approx 10^4$. For larger values of $N_k$, we approximate $P(\mu_k \vert \theta_k)$ as a Gaussian distribution centered in 
$\mean{\J_z^{\rm out}}$ and width $(\Delta \J_z^{\rm out})^2$, where $\J_z^{\rm out}$ can be expressed as a function of the spin moments of the states after 
state-preparation as $\J_z^{\rm out}  = \J_x \sin \theta_k + \J_z  \cos \theta_k$.
We have checked the equivalence of the two approaches for small values of $N_k$.\\

{\bf Acknowledgments.}
%We acknowledge funding from the project EMPIR-USOQS, EMPIR projects are co-funded by the European Unions Horizon2020 research and innovation programme and the EMPIR Participating States. 
We acknowledge financial support from the European Union’s Horizon 2020 research and innovation programme - Qombs Project, FET Flagship on Quantum Technologies grant no. 820419.
%, and from the H2020 QuantERA ERA-NET Cofund in Quantum Technologies projects QCLOCKS and CEBBEC.

\end{document}